\documentstyle[aps,twocolumn, epsfig]{revtex}

\begin{document}
\draft
\date{\today}

\title{M\"ossbauer effect for dark solitons in Bose-Einstein condensates}

\author{Th. Busch and J. R. Anglin}

\address{Institut f\"ur Theoretische Physik, 
Universit\"at Innsbruck, 
A--6020 Innsbruck, AUSTRIA}

\maketitle

\begin{abstract}
We show that the energetic instability of dark solitons is associated with
particle-like motion, and present a simple equation of
motion, based on the M\"ossbauer effect,
for dark solitons propagating in inhomogeneous Thomas-Fermi clouds. Numerical
simulations support our theory. We discuss some experimental approaches.
\end{abstract}


\date{\today}

\narrowtext

The success of the Gross-Pitaevski mean field theory in describing 
experimentally observed dilute
Bose condensates\cite{Science95,Bradley,Davis} shows that one really can 
persuade a large number of particles
to behave as a field.  The perversely obvious thing to do with such a field,
then, is to persuade it to behave as a particle.  This would not only be a
pleasant closing of a circle, but would link condensate physics to other 
subjects in which solitons are important, such as fluid mechanics, nonlinear 
optics, and fundamental particle theory.  Perhaps most importantly, it might
offer a powerful laboratory for studying the interface between quantum and
classical physics, by providing a wholly artificial classical particle, 
composed of
highly controllable, weakly interacting quantum particles.  In this Letter
we discuss one particular particle-like configuration of the Gross-Pitaevski
mean field, namely the one-dimensional {\it dark soliton}.  We describe the
behaviour of dark solitons in inhomogeneous potentials, reporting both
analytical and numerical results, and we propose some experimental approaches
for effectively one-dimensional traps.  We also suggest that the dark soliton 
should be considered the simplest member of a family of technically unstable 
but nevertheless robust monopoles in $D$ dimensions. 

The Gross-Pitaevski equation (GPE) governs the evolution of the c-number 
`macroscopic wave function' $\psi(\vec{x},t)$ of a Bose-Einstein condensate.
By appropriately scaling the wave function, space, and time, and incorporating
a chemical potential by extracting a factor $e^{-i\kappa^2 t}$, one can write
this equation in the convenient form   
\begin{equation}\label{GPE} 
i\partial_t\psi = -{1\over2}\nabla^2\psi 
        +(|\psi|^2 +V(\vec{x})-\kappa^2)\psi\;. 
\end{equation} 
We have assumed a positive scattering length, and have not restricted the 
normalization constant $\int\!dx\,|\psi|^2$.  We will focus for most of
this paper on the limit of a long trap so thin that one can apply the GPE in 
one dimension.  The approach to this limit from three dimensions has recently
been discussed\cite{Pethick}.  

Eqn.~(\ref{GPE}) in one dimension with $V=0$ has been extensively 
studied in nonlinear optics\cite{DSrep}, and a particle-like solution
has long been known: $\psi_{DS} = \kappa\tanh(\kappa x)$.
This time-independent solution is known as a  {\it dark soliton}, because it 
describes a small dark spot in a light pulse.  
If $\psi(x)$ were restricted to be real, the dark soliton would
be topologically stable in the same way as the Sine-Gordon kink
that it strongly resembles.  Since $\psi$ is very nearly constant far from 
$x=0$, but changes sign within a few healing lengths, it resembles a domain
wall: a one-dimensional monopole.

There exist close relatives of the dark soliton which may be recognized as
$D$-dimensional monopoles; we digress briefly to sketch the family group.
If we allow $\psi$ to be a multi-component object $\psi =
(\psi_1, ..., \psi_D)$, and interpret
$|\psi|^2=\sum_j|\psi_j|^2$, then (\ref{GPE}) is the mean field
equation of motion for a condensate of atoms with $D$ internal
states (among which all scattering lengths are equal).  In two 
dimensions one finds $\psi = f(\kappa r)
(\cos\theta,\sin\theta)$ to be a solution to 
the two component GPE, where $f$ is the radial profile of a one-component 
vortex.  Analytical and numerical results related to this solution will be 
reported 
elsewhere.  The generalization to three dimensions for three components is 
obvious\cite{gyropole}. In all these cases,
if $\psi$ were restricted to be real we would have topologically stable
monopoles, as small `bubbles' in the condensate.
  
It is not difficult to see,
however, that by taking $\psi$ into the complex plane one can
continuously deform it into a configuration with constant
density, eliminating the `bubble' in $|\psi|^2$. 
One can exhibit this instability more precisely by exploring the
free energy of nearby configurations; although this analysis can be carried
out more generally, we now revert to one dimension.  As a time-independent
solution to (\ref{GPE}) ($V=0$, one dimension), 
$\psi_{DS}=\kappa\tanh\kappa x$ is a 
stationary point of
the free energy functional 
\begin{equation}\label{G}
G={1\over2}\int\!dx\,\big[|\psi'|^2+(|\psi|^2-\kappa^2)^2\bigr]\;. 
\end{equation} 
One can diagonalize the Hessian matrix of $G$ at
the stationary point by solving what may be recognized as
imaginary time Bogoliubov equations for the perturbation
$\delta\psi$.  
(The real time Bogoliubov theory also diagonalizes the
Hessian matrix, but by a symplectic instead of orthogonal transformation,
as required by canonicity.)
Because $\psi$ is real, the real and imaginary
parts of $\delta\psi=R+iS$ decouple, and independently satisfy
(different) Schr\"odinger equations: the normal modes $R_k(x)$ and $S_k(x)$ are
respectively eigenstates of $\hat{H}_3$ and $\hat{H}_1$, for
\begin{equation}\label{IBOG}
\hat{H}_n \equiv -{1\over2}{d^2\over dx^2} -\kappa^2 + n\psi_{DS}^2\;.
\end{equation} 
One finds the `core bound
state' $S_0 = 1/\cosh\kappa x$, for which the Hessian matrix
has a negative eigenvalue; this shows that there are
configurations near the dark soliton with lower free energy. 
And in numerical evolution of the dark soliton in imaginary
time, numerical noise soon finds this
unstable mode, and fills in the monopole bubble.

Imaginary time evolution has only the most indirect implications
for real time evolution, however.  In one dimension with $V=0$
constant, we have the following two parameter family of exact,
but not generally time-independent, solutions to (\ref{GPE}):
\begin{eqnarray}\label{grey}
\psi_{q,\dot{q}}(x,t) &=& i\dot{q} + 
        \sqrt{\kappa^2-\dot{q}^2}\tanh\sqrt{\kappa^2-\dot{q}^2}(x-q)\;, 
\end{eqnarray}
where $\dot{q}$ is the time derivative of $q$, constant and satisfying 
satisfying $|\dot{q}|<\kappa$.

For these moving monopoles (familiar in nonlinear optics as `grey solitons'), 
$|\psi|^2$ never
drops below $\dot{q}^2$; and the phase slip across the soliton is $\pi
-\arctan (|\dot{q}|/\sqrt{\kappa^2-\dot{q}^2})$.  This means that in the limit
$\dot{q}\to\pm\kappa$, where the soliton is travelling at the
superfluid critical velocity, the dark soliton actually vanishes, becoming
identical with motionless condensate.  But the free energy of
the grey soliton is 
\begin{equation}\label{Ggrey} G(q,\dot{q}) = 
{4\over3}(\kappa^2-\dot{q}^2)^{3\over2}\;.  
\end{equation} 
This implies that {\it grey solitons have negative kinetic energy.}  
The imaginary time Bogoliubov equations found
only one negative energy degree of freedom near the dark
soliton, and we can now identify it in the real time Bogoliubov theory
as the canonical momentum conjugate to the translational zero mode.  
And so we see that {\it the instability of the dark soliton is not merely to
break up or fill in, but to acquire velocity}.  

Since dark solitons have both translational modes and
independent velocities, they have the phase space of a particle. 
This is in contrast to a vortex in two dimensions, whose two
positional degrees of freedom are in fact canonically conjugate,
so that the vortex phase space is only two
dimensional\cite{vortexmotion}. It is therefore natural to seek
a particle-like second order equation of motion for dark solitons
in an inhomogeneous potential $V(x)$.  One can readily guess that,
at least in a sufficiently slowly varying potential, one should simply replace
$\kappa^2\to\kappa^2-V(q)$ in (\ref{grey}), and insert the result into a
background Thomas-Fermi cloud.  The question of how such a configuration
will evolve in time, however, is not trivial.  Time-dependent variational 
approaches
can be informative but also misleading.  Fortunately, a more systematic 
procedure exists.

In the experimentally relevant situation  
where $V(x)$ varies slowly on the healing length scale $|\psi|^{-1}$, one 
can apply a time-dependent boundary layer theory similar to that used
for vortices in Ref.~\cite{vortexmotion}. 
Perturb around the Thomas-Fermi
approximation everywhere outside a small moving `core zone' extending a
few healing lengths from the monopole; treat the inhomogeneity
of $V$ as a perturbation within the core zone; and then match
solutions smoothly at the moving boundary.  This procedure 
leads to a very simple result in the case where the only excitations
present (other than the soliton motion) are those generated by the soliton as 
it accelerates or decelerates.  In this case the boundary layer theory 
demonstrates
that when the healing length is much smaller than the Thomas-Fermi cloud, the
system is in the M\"ossbauer limit, in which phonon mode excitations conserve
momentum without contributing significantly to the total energy.  As a result,
the soliton energy $[\kappa^2-\dot{q}^2-V(q)]^{3/2}$ is conserved to leading
order in the ratio of length scales, and this implies the 
equation of motion\cite{clarkcom}
\begin{equation}\label{EofM}
\ddot q = - {1\over2}V'(q)\;.
\end{equation}

In a harmonic trap, Eqn.~(\ref{EofM}) implies oscillation of the 
soliton\cite{footnoteHong} 
with frequency $1/\sqrt{2}$ times that of the dipole mode of the condensate.  
We have confirmed the frequency factor to 
rather more than the expected accuracy in 
numerical simulations of harmonic traps over a wide range of condensate 
densities and 
oscillation amplitudes; we have also monitored the center of mass motion, and 
confirmed that it is decoupled and has the trap frequency.  Eqn.~(\ref{EofM}) 
also holds for arbitrary potentials, however, as long as they vary slowly on 
the healing 
length scale.  We have therefore further confirmed the good accuracy of our 
M\"ossbauer limit equation of motion by solving 
Eqn.~(\ref{GPE}) numerically over a wide range of parameters and for 
various potentials; a generic sample is shown in Fig.~1.

To solve the NLSE numerically we use the split-step Fourier method
\cite{splitstep}, in two stages.  First we construct an initial state with no
excitations apart from the soliton, by propagating a crude trial wavefunction 
in imaginary time, and re-normalizing it between each step. Eventually, 
numerical noise excites the `core bound state' and the monopole rapidly 
disappears in imaginary time; but before this there is a clear plateau in 
energy, during which the monopole slowly drifts in the Thomas-Fermi
cloud, in the direction of decreasing density.  We simply stop the imaginary
time evolution during this period, and save the configuration.  Then we 
propagate this optimized initial state in real time.  As is well known, real 
time evolution is prone to (clearly recognizable) high frequency numerical 
instabilities; but by keeping our time steps
shorter than $h^2/\pi$, where $h$ is the spatial grid interval, we are able to 
simulate several periods of monopole
motion, even at condensate densities approaching those attained in real 
experiments.  (At $\int\!dx\,|\psi|^2=1000$, our 
Thomas-Fermi cloud extends to over ten times the trap width.)

Since Eqn.~(\ref{EofM}) predicts a particular frequency for finite amplitude 
oscillations around a local minimum of the trap potential, and a dynamical 
instability for motion near a local maximum, it would be interesting to 
compare these predictions with those from Bogoliubov theory. Since we have
numerically found frequencies very close to that from Eqn.~(\ref{EofM}) even 
for extremely small amplitude oscillations, we strongly suspect
that the two theories agree in the M\"ossbauer limit.  
Such agreement is not actually necessary, 
however: Eqn.~(\ref{EofM}) is derived for arbitrary 
amplitude oscillations but neglecting coupling to phonon modes, while
the Bogoliubov spectrum governs only infinitesimal perturbations, but
includes all excitations.  Co-incidence of the two spectra would imply that the
dark soliton decouples from phonon modes at extrema of the potential, a result
which might be of some importance for the problems of dark soliton zero point 
motion and mesoscopic tunneling.

\begin{center}
 \begin{figure}
 \epsfig{file=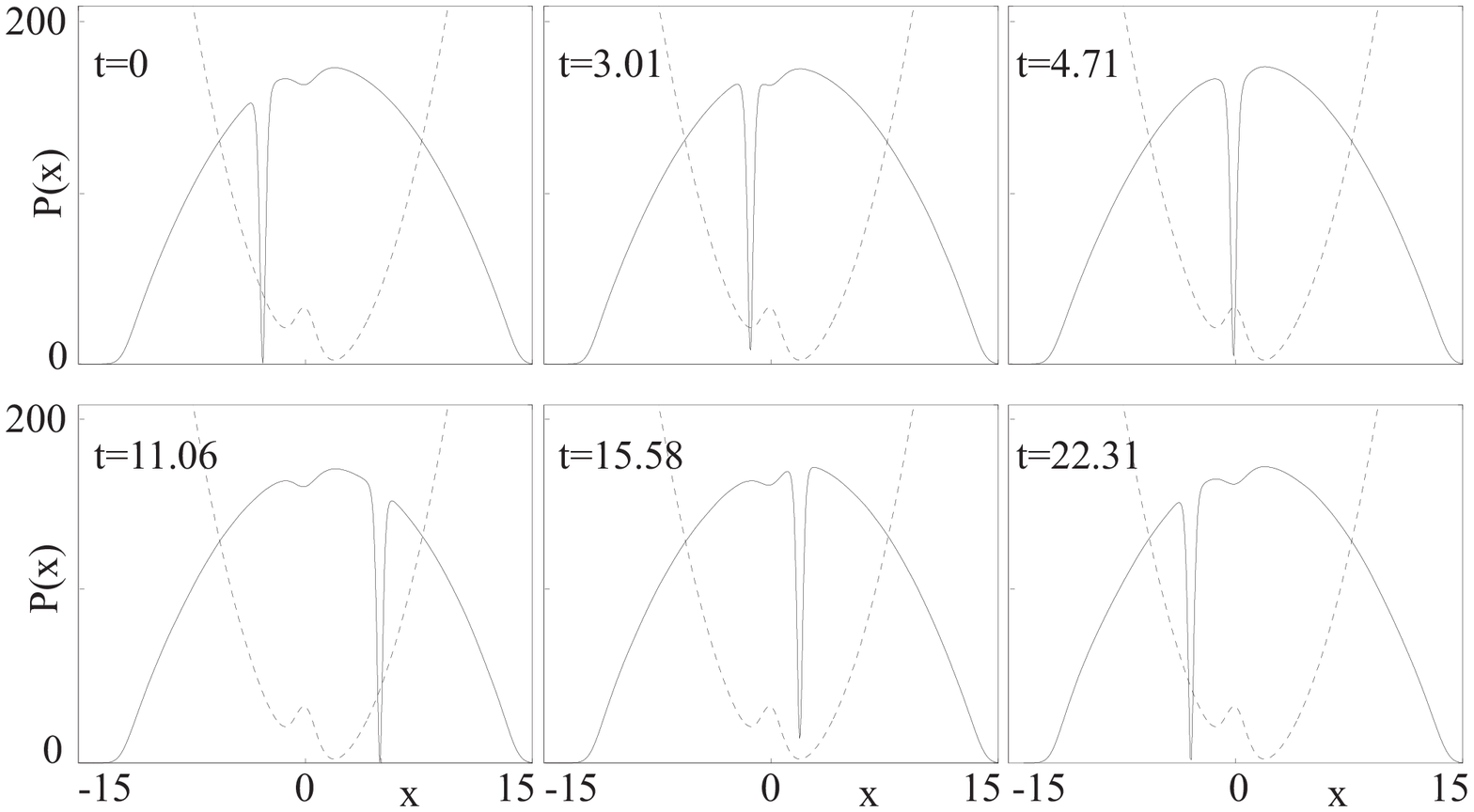,width=\linewidth}
 \caption{Normalized $|\psi|^2$ ($\int\!dx\,|\psi|^2=300$) for a dark soliton 
oscillating in an asymmetric potential with a bump, shown (magnified) in dots. 
The potential is $V=0.1 x(x-2)+1.1/\cosh^2(x)$, the soliton starting point 
is x=-2.81 
(with a grid uncertainty of about 0.03); Eqn.~(\ref{EofM}) predicts the 
interval between successive times of complete darkness ($|\psi|^2=0$ at the
minimum) to be $T/2=11.5\pm0.1$.  Comparing times in lower left and lower right
plots shows the M\"ossbauer limit is good to within 5\%; the ratio of healing
length $1/|\psi|$ to potential scale is of this order.}
 \end{figure}
\end{center}

Since the equation of motion (\ref{EofM}) is obtained by neglecting excitation
of phonon modes by the monopole motion, but in fact some excitation does occur,
we must expect dissipation (which could in principle be calculated by 
continuing the time-dependent boundary layer theory to higher orders in the
ratio of length scales).  Since the soliton energy is negative, however,
dissipation has the effect of {\it increasing} the amplitude of the monopole 
motion in the trap. We have observed this effect in numerical simulations:
phonon mode excitations are visible in slight distortions of the Thomas-Fermi
envelope.  These excitations produce small variations, plus a slow trend 
upwards, in $\dot{q}^2+V(q)$; see Fig.~2.  

\begin{center}
 \begin{figure}
 \epsfig{file=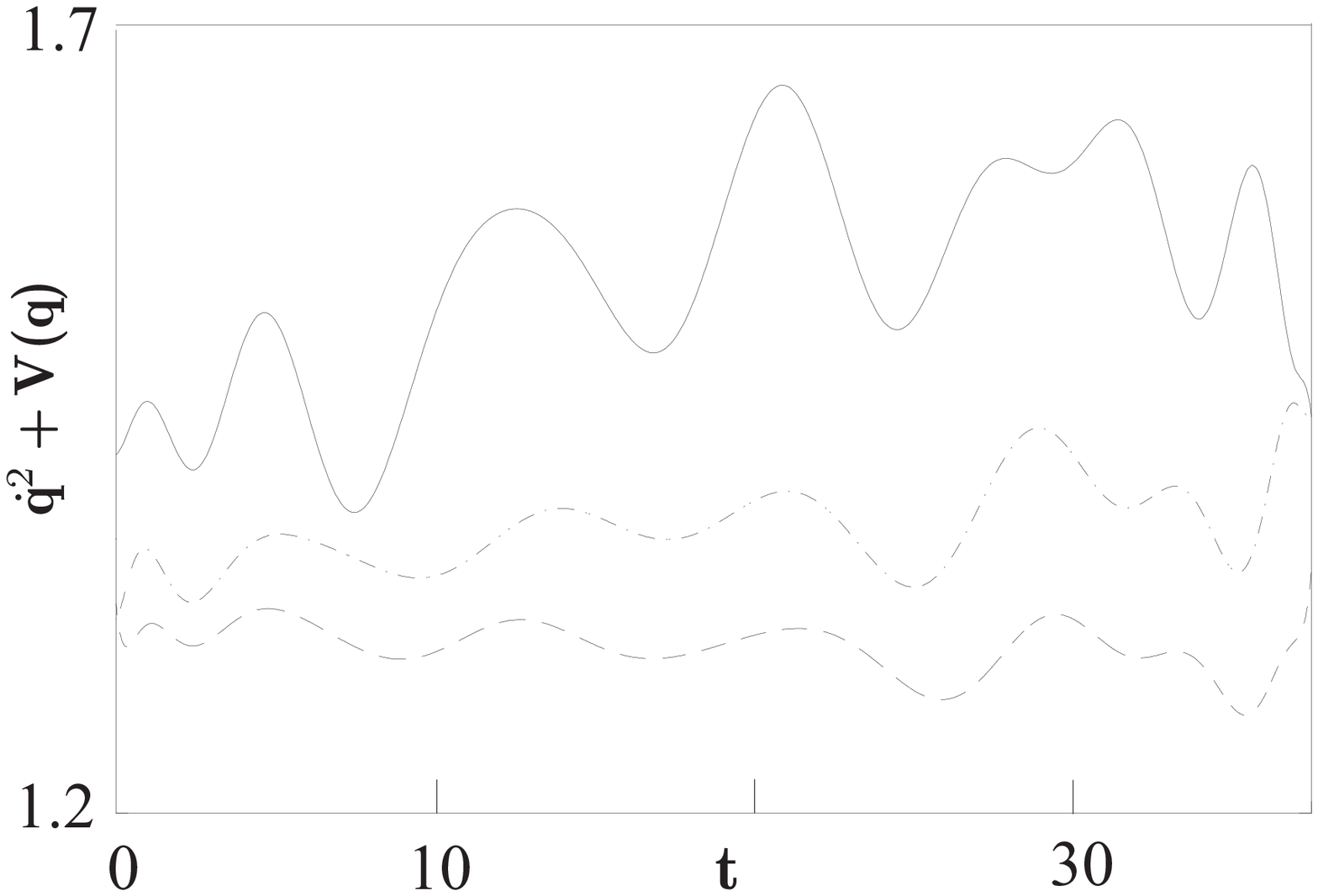,width=.9\linewidth, height=4.5cm}
 \caption{Plots of the quantity $\dot{q}^2+V(q)$, conserved in the 
M\"ossbauer limit, versus time, as a dark soliton moves in the potential of 
Fig.~1. The three curves are for various densities: 
$\int\!dx\,|\psi|^2=100,500,1000$ from top to bottom. For $\dot{q}^2$ we 
actually take the height of $|\psi|^2$ at the soliton minimum, which is equal 
to $\dot{q}^2$ for dark solitons in bulk.}
 \end{figure}
\end{center}

We conclude our theoretical discusion by remarking on the interaction of 
two solitons.
Analytical work on dark solitons in bulk has shown that two dark solitons
interact through a short range repulsive potential, whose maximum height is
finite and velocity dependent\cite{DSint}.  
So solitons more than a few healing lengths apart do not influence
one another, and solitons colliding rapidly can pass through each other with
negligible interaction; but two slow solitons cannot pass each other.  
Moreover, the transition between impenetrability and near-freedom is 
fairly sharp as a function of relative velocity.  On the other hand, the
presence of a second soliton in a harmonic trap appears to {\it shorten} the 
oscillation period of the pair; see Fig.~3.  

\begin{center}
 \begin{figure}
 \epsfig{file=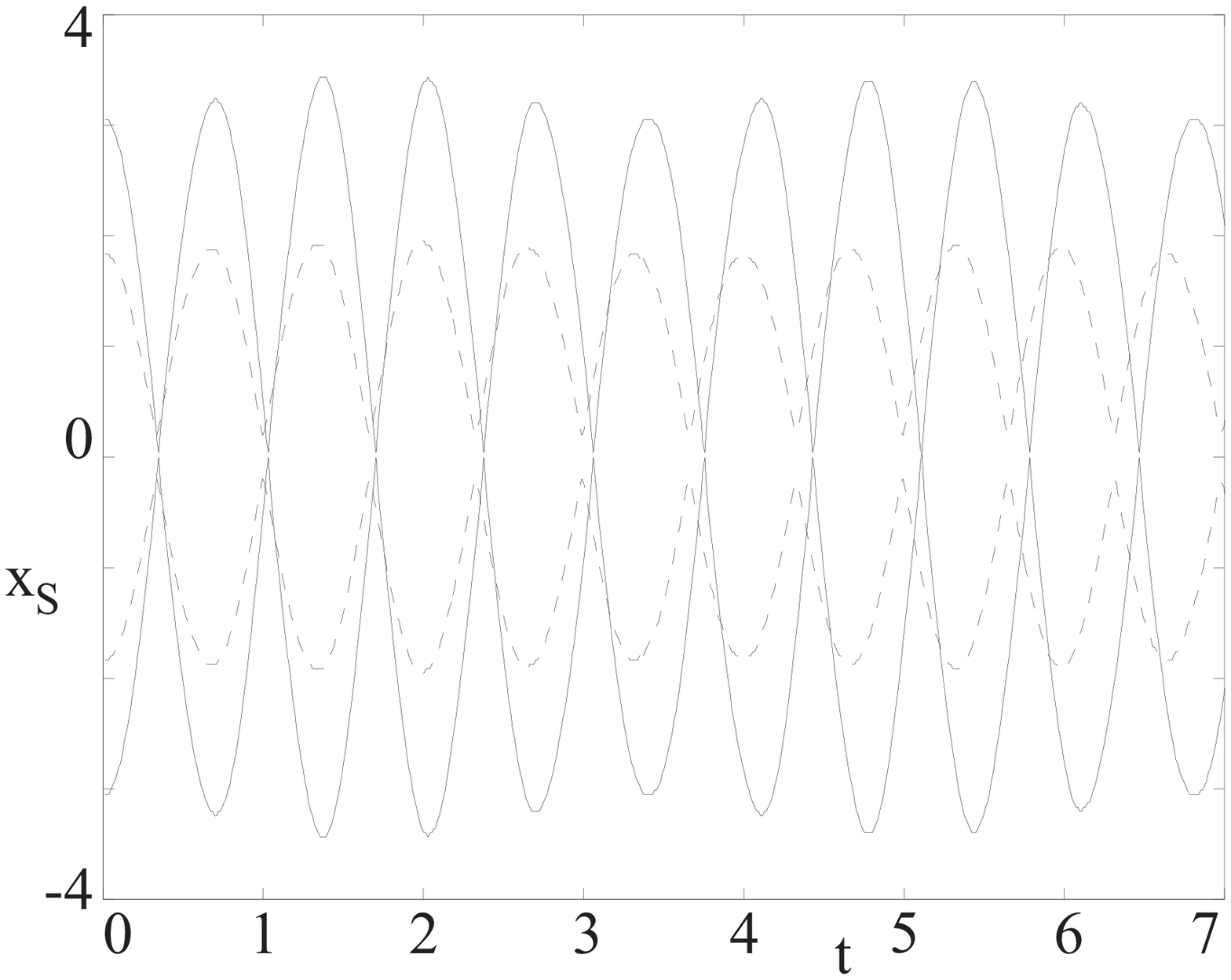,width=.9\linewidth}
 \caption{Two different cases of two solitons oppositely displaced in a 
harmonic trap.  The trap angular frequency is one, and the time axis is shown 
in units of $2\pi$.  The solid line shows solitons passing through each other;
the dashed line, for two solitons initially displaced somewhat less, shows 
bouncing motion.  For both cases, $\int\!dx\,|\psi|^2=100$.  Note that the 
period of bouncing motion is approximately $2\pi\times 1.33$, that of passing
motion $2\pi\times 1.36$.  For comparison, the period of a single soliton at 
this density is $2\pi\times 1.42$.  Both motions are also clearly modulated, 
due to coupling to a phonon mode.}
 \end{figure}
\end{center}

We now briefly outline experimental approaches to the creation of dark 
solitons.
An ingenious proposal has recently been presented to construct dark solitons by
adiabatic passage\cite{Dum}.  This scheme requires that the constructed soliton
be a dynamically stable stationary state, which would seem to prevent one from
making dark solitons on slopes or local maxima of the potential.  Since the
capability exists to make temporary small local wells anywhere within the 
larger trap, however, this need not actually pose a problem.  (This capability
should also allow more general testing of the dark soliton equation of motion, 
as one could arrange a sequence of `speed bumps', or accelerating wells, along
the thin trap.)  Other recent work
has noted that dark solitons are generically produced during collisions between
condensates\cite{Clark,Burnett}, or between generic travelling 
disturbances\cite{Huo}.

Another challenge is to detect dark solitons once they are made, since they
are typically much smaller than a wavelength of light.  Current techniques of
relaxing the trap so that the soliton `hole' grows to visibility should be
helpful, as should the interferometric result that a dark soliton produces a 
discontinuity in the pattern of Ramsay fringes.  We would like to suggest an
alternative method, however, which could allow one to view dark solitons 
directly and non-destructively: trapping atoms with trapped atoms.  Since 
there is a bound state in the dark 
soliton potential, a few atoms of another species could in principle be bound
inside the dark soliton as it moves (assuming that they are repelled by the 
condensate atoms).  In principle one might hope to introduce these probe atoms
simply by `dusting' the condensate with many of them, so that most probe 
atoms are simply 
repelled into the surrounding thermal cloud, but some are sympathetically 
cooled and `stick' inside any dark solitons which may exist. Once there, they
may be made to fluoresce independently of the condensate, so that the `dark' 
solitons may be followed as small bright spots. Further investigation of this
proposal is obviously required, as some condensate atoms are lost as the probe 
atoms are cooled sympathetically, and one must ensure that the capture cross
section of the soliton is high enough that one does not need so much `dust' 
that the soliton, or even the condensate, is destroyed.  In principle however 
it seems a promising way to observe many types of topological defects.

As a final experimental remark, we note that in traps which are not 
sufficiently close to the one-dimensional limit one expects dark solitons to be
unstable, presumably above some threshold velocity, to the formation of 
vortex rings; these would of course be at least as interesting as pure 
solitons.

In conclusion we wish to re-emphasize that although dark solitons are unstable,
this instability should be much more of an advantage than a problem, as it 
allows the solitons to behave as particles, with independent position and 
velocity.
In the M\"ossbauer limit a very simple equation of motion applies, which is
expected to remain valid while the amplitude of soliton motion in the 
potential slowly
grows.  The final victory of the instability is the escape of the dark soliton
from the trap: at least for these mesoscopic `hollow men', the world will
after all end with more of a bang than a whimper\cite{Eliot}.

\begin{center}{\bf Acknowledgements}\end{center}

We are happy to acknowledge valuable discussions with Ignacio Cirac, Victor
Perez-Garcia, Peter Zoller, and Wojciech Zurek.  JRA is grateful for the 
hospitality of the
Benasque Centre for Theoretical Physics, where part of this work was done.  
This work was supported by the European Union under the TMR 
Network ERBFMRX-CT96-0002 and by the Austrian Fond zur F\"orderung der 
wissenschaftlichen Forschung.



\begin{references}


\bibitem{Science95}
M. H. Anderson, J. R. Ensher, M. R. Matthews, C. E. Wieman, E. A. Cornell
Science, {\bf 269}, 198 (1995).

\bibitem{Bradley} 
C. C. Bradley, C. A. Sackett, J. J. 
Tollett, R. G. Hulet, Phys. Rev. Lett. {\bf 75}, 1687 (1995).

\bibitem{Davis} 
K. B. Davis, M.-O. Mewes, M. R. Andrews, N. J. van Druten, D. S. Durfee, 
D. M. Kurn, W. Ketterle, Phys. Rev. Lett. {\bf 75}, 3969 (1995). 

\bibitem{Pethick}
A.~D.~Jackson, G.~M.~Kavoulakis, C.~J.~Pethick, Phys. Rev. A  {\bf 58}, 2417 
(1998) 

\bibitem{DSrep} Y.S. Kivshar and B. Luther-Davies, Phys. Rep. {\bf 298}, 81 
(1998). 

\bibitem{gyropole} It is not possible to make a two-component
monopole by wrapping the Bloch sphere once into three-space,
because the resulting $\psi$ is not single-valued. But W.H.
Zurek (private communication) has suggested what does seem to be
a viable two-component three-dimensional monopole: $\psi =
g(\kappa r)(\cos\theta, e^{i\phi}\sin\theta)$.  We suggest the name
`gyropole' for this configuration, which is a
kind of hybrid between a two-dimensional vortex and a dark
soliton.


\bibitem{vortexmotion}
 B.~Y.~Rubinstein, L.~M.~Pismen, Physica D {\bf 78}, 1 (1994)

\bibitem{clarkcom} A similar equation was presented for dark solitons 
in Ref.\cite{Clark}, but (when their equation
is translated into our units) without the factor of $1/2$.  That result 
actually applies only
to bright solitons, which exist for negative scattering lengths; since 
bright solitons are isolated micro-condensates, their equation of motion 
follows easily from Ehrenfest's theorem.

\bibitem{footnoteHong} Oscillatory motion is alluded to in a brief reference 
to unpublished work in Ref.\cite{Huo}, and
it is also of course qualitatively predicted by the semi-classical equation of
Ref.\cite{Clark}.

\bibitem{splitstep}
J.~A.~C.~Weideman, B.~M.~Herbst, SIAM J. Numer. Anal. {\bf 23}, 485 (1986).

\bibitem{DSint} Y.S. Kivshar and W. Kr\'olikowski, Opt. Comm. {\bf 114}, 353 
(1995), and references therein.

\bibitem{Dum}
R. Dum, J.I. Cirac, M. Lewenstein, P. Zoller, Phys. Rev. Lett.
{\bf 80}, 2972 (1998) 

\bibitem{Clark} 
W.~P.~Reinhardt, C.~W.~Clark, J. Phys. B {\bf 30}, L785 (1997)

\bibitem{Burnett}
T.~F.~Scott, R.~J.~Ballagh, K.~Burnett, cond-mat/9711111

\bibitem{Huo}
T.~Hong, Y.~Z.~Wang, Y.~S.~Huo, Phys. Rev. A  {\bf 58}, 3128 (1998) 

\bibitem{Eliot} T.S. Eliot, `The Hollow Men', 
in {\it Collected Poems 1909-1962} (Faber and Faber, 1963).


\end{references}
\end{document}